\documentclass[usegraphicx,usenatbib,useAMS,twocolumn]{mn2e}

\usepackage{epsfig}
\usepackage{graphicx}
\usepackage{amsmath,bm}
\usepackage{color}
\usepackage{amssymb}
\usepackage{times}
\usepackage{multirow}

\title[BAO measurement from the LOS-dependent power spectrum of DR12 BOSS galaxies]{The clustering of galaxies in the SDSS-III Baryon Oscillation Spectroscopic Survey: BAO measurement from the LOS-dependent power spectrum of DR12 BOSS galaxies}

\author[H. Gil-Mar\'in et al.]{H\'ector Gil-Mar\'in$^{1}$\thanks{hector.gil@port.ac.uk},  Will J. Percival$^{1}$, Antonio J. Cuesta$^{2}$, Joel R. Brownstein$^{3}$, \and Chia-Hsun Chuang$^{4}$, Shirley Ho$^{5}$, Francisco-Shu Kitaura$^{6}$, Claudia Maraston$^1$, \and Francisco Prada$^{4,7,8}$, Sergio Rodr{\'i}guez-Torres$^{4,7,9}$, Ashley J. Ross$^{10}$, \and David J. Schlegel$^{11}$, Donald P. Schneider$^{12,13}$, Daniel Thomas$^1$, Jeremy L. Tinker$^{14}$,\and Rita Tojeiro$^{15}$, Mariana Vargas Maga\~na$^{16}$ \& Gong-Bo Zhao$^{1,17}$ \\
 $^{1}$ Institute of Cosmology \& Gravitation, University of Portsmouth, Dennis Sciama Building, Portsmouth PO1 3FX, UK\\
 $^{2}$ Institut de Ci\`encies del Cosmos, Universitat de Barcelona, IEEC-UB, Mart\'i i Franqu\`es 1, 08028, Barcelona, Spain\\
 $^{3}$ Department of Physics and Astronomy, University of Utah, 115 S 1400 E, Salt Lake City, UT 84112, USA\\
   $^4$ Instituto de F\'{\i}sica Te\'orica, (UAM/CSIC), Universidad Aut\'onoma de Madrid,  Cantoblanco, E-28049 Madrid, Spain \\
       $^{5}$ McWilliams Center, Carnegie Mellon University, Pittsburgh, PA 15213, USA \\
       $^{6}$ Leibniz-Institut f{\"u}r Astrophysik (AIP), An der Sternwarte 16, D-14482 Potsdam, Germany \\
       $^{7}$ Departamento de F\'{\i}sica Te\'orica, Universidad Aut\'onoma de Madrid, Cantoblanco, 28049, Madrid, Spain \\
$^{8}$ Instituto de Astrof\'{\i}sica de Andaluc\'{\i}a (CSIC), Glorieta de la Astronom\'{\i}a, E-18080 Granada, Spain \\ 
$^9$ Campus of International Excellence UAM+CSIC, Cantoblanco, E-28049 Madrid, Spain \\
 $^{10}$ Center for Cosmology and AstroParticle Physics, The Ohio State University, Columbus, OH 43210, USA\\
 $^{11}$ Lawrence Berkeley National Lab 1 Cyclotron Road, Berkeley, CA 94720 USA \\
 $^{12}$ Department of Astronomy and Astrophysics, The Pennsylvania State University, University Park, PA 16802, USA\\
$^{13}$  Institute for Gravitation and the Cosmos, The Pennsylvania State University,  University Park, PA 16802, USA\\
$^{14}$ Center for Cosmology and Particle Physics, Department of Physics, New York University, New York, NY 10003, USA\\
$^{15}$ School of Physics and Astronomy, St Andrews KY16 9SS, UK\\
$^{16}$ Instituto de Fisica, Universidad Nacional Aut\'onoma de M\'exico, Apdo. Postal 20-364, M\'exico  \\
$^{17}$National Astronomy Observatories, Chinese Academy of Science, Beijing, 100012, P. R. China }
\def\gs{\mathrel{\raise1.16pt\hbox{$>$}\kern-7.0pt
\lower3.06pt\hbox{{$\scriptstyle \sim$}}}}         
\def\ls{\mathrel{\raise1.16pt\hbox{$<$}\kern-7.0pt 
\lower3.06pt\hbox{{$\scriptstyle \sim$}}}}         

\begin{document}
\maketitle

\begin{abstract} 
We present an anisotropic analysis of the baryonic acoustic oscillation (BAO) scale in the twelfth and final data release of the Baryonic Oscillation Spectroscopic Survey (BOSS). We independently analyse the LOWZ and CMASS galaxy samples: the LOWZ sample contains contains 361\,762 galaxies with an effective redshift of $z_{\rm LOWZ}=0.32$; the CMASS sample consists of 777\,202 galaxies with an effective redshift of $z_{\rm CMASS}=0.57$. We extract the BAO peak position from the monopole power spectrum moment, $\alpha_0$, and from the $\mu^2$ moment, $\alpha_2$, where $\mu$ is the cosine of the angle to the line-of-sight. The $\mu^2$-moment provides equivalent information to that available in the quadrupole but is simpler to analyse. After applying a reconstruction algorithm to reduce the BAO suppression by bulk motions, we measure the BAO peak position in the monopole and $\mu^2$-moment, which are related to radial and angular shifts in scale.  We report  $H(z_{\rm LOWZ})r_s(z_d)=(11.60\pm0.60)\cdot10^3\,{\rm km}s^{-1}$ and $D_A(z_{\rm LOWZ})/r_s(z_d)=6.66\pm0.16$ with a cross-correlation coefficient of $r_{HD_A}=0.41$, for the LOWZ sample; and $H(z_{\rm CMASS})r_s(z_d)=(14.56\pm0.37)\cdot10^3\,{\rm km}s^{-1}$ and $D_A(z_{\rm CMASS})/r_s(z_d)=9.42\pm0.13$ with a cross-correlation coefficient of $r_{HD_A}=0.47$, for the CMASS sample. We demonstrate that our results are not affected by the fiducial cosmology assumed for the analysis. We combine these results with the measurements of the BAO peak position in the monopole and quadrupole correlation function of the same dataset \citep[][companion paper]{Cuestaetal2015} and report the consensus values: $H(z_{\rm LOWZ})r_s(z_d)=(11.63\pm0.69)\cdot10^3\,{\rm km}s^{-1}$ and $D_A(z_{\rm LOWZ})/r_s(z_d)=6.67\pm0.15$ with $r_{HD_A}=0.35$ for the LOWZ sample; $H(z_{\rm CMASS})r_s(z_d)=(14.67\pm0.42)\cdot10^3\,{\rm km}s^{-1}$ and $D_A(z_{\rm CMASS})/r_s(z_d)=9.47\pm0.12$ with $r_{HD_A}=0.52$ for the CMASS sample.
  \end{abstract}


\section{Introduction}\label{sec:intro}

The Baryon Acoustic Oscillation (BAO) signal in the clustering of galaxies provides a robust route to measure the cosmological expansion rate \citep{Eis05}. The angular separation of galaxies conveys different cosmological information compared to the radial separation. The observed projection of the BAO-scale in the angular direction depends on the angular diameter distance, while the radial projection is a function of the Hubble expansion\footnote{Negligible evolution is expected over the redshift interval between galaxies separated by the BAO scale}. The anisotropic information can be extracted from measuring the BAO peak position in moments of the correlation function or Fourier-space power spectrum. 

The Baryon Oscillation Spectroscopic Survey  (BOSS; \citealt{Dawsonetal:2012}), which is part of SDSS-III \citep{EISetal:2011}, has provided the largest set of numbers of spectroscopic galaxy observations made to date; in this paper we analyse the final, Data Release 12 (DR12; \citealt{dr12}) sample. Our work follows on from measurements made for DR11 samples presented in \citet{Andersonetal:2014} and \citet{Beutleretal:2013}. We  analyse the low-redshift (LOWZ) and high-redshift (CMASS) samples independently, which were targeted using different algorithms \citep{Reidetal:2015}. DR12 contains approximately 1000\,deg$^2$ more solid angle than DR11, an increase of approximately 12\%. Additionally, the methodology adopted to create galaxy catalogues has been improved \citep{Reidetal:2015, Ross15}, leading to improved understanding, and mitigation, of potential systematic errors. 

In this paper we present a Fourier-space analysis of the line-of-sight (LOS) dependent clustering, extracting the BAO position. The monopole and quadrupole calculated were analysed to measure the Redshift-Space Distortion (RSD) signal in \citet{paperRSD},  hereafter Paper I. In this paper we present a new method for fitting the $\mu^2$-moment of the power spectrum rather than the quadrupole to obtain the anisotropic information. As the $\mu^2$-moment is a simple linear combination of monopole and quadrupole, this change is lossless. Fitting to the $\mu^2$-moment allows the use of the same equations as for the monopole and quadrupole (with different parameters) simplifying the modelling, and particularly the way in which we can isolate BAO information from the broadband signal.  Note that using the $\mu^2$-moment instead of the quadrupole is a mere fitting technique that simplifies the identification of the BAO peak in the higher order multipoles of the power spectrum. However, one can still use the estimators used in Paper I to compute the monopole and quadrupole, and reconvert them into monopole and $\mu^2$-moment. Since the $\mu^2$-moment is just a linear combination of the monopole and quadrupole, the result is identically the same to measure the $\mu^2$-moment using the estimator described in \cite{Bianchietal:2015}. 

The outline of our paper is as follows: In \S\ref{sec:catalogs}, we introduce the data, the mocks used to compute the covariance matrices and test for systematics, and briefly describe the reconstruction technique. In \S\ref{sec:estimator_ps} we present the formalism used to calculate the monopole and $\mu^2$-moments. The method applied to fit the BAO is described in Section~\ref{sec:methodology}. The measurements of the BAO peak position in the DR12 BOSS galaxies are presented in \S\ref{sec:results}, for both pre- and post-reconstructed catalogues. In \S\ref{sec:test_mocks} we test for potential systematics of the model and of the fitting process using the post-reconstructed mocks. Finally  \S\ref{sec:conclusions} presents the conclusions of this paper. 

\section{Data and Mocks}\label{sec:catalogs}
\subsection{The SDSS III BOSS data}\label{sec:bossdata}

As part of the Sloan Digital Sky Survey III \citep{EISetal:2011} the Baryon Oscillations Spectroscopic Survey (BOSS) \citep{Dawsonetal:2012}  measured spectroscopic redshifts \citep{Bolton12,Smee13} for more than 1 million galaxies and over 200\,000 quasars. The galaxies were selected from multi-colour SDSS imaging \citep{Fukugitaetal:1996,Gunnetal:1998,Smithetal:2002,Gunnetal:2006,Doietal:2010} focussing on the redshift range of $0.15\leq z \leq0.70$. The galaxy survey used two primary target algorithms, selecting samples called LOWZ, with 361\,762 galaxies in the final data release DR12 \citep{dr12} between $0.15\leq z \leq0.43$ and CMASS, with 777\,202 galaxies in DR12 between $0.43\leq z \leq 0.70$. The full targeting algorithms used and the method for calculating the galaxy and random catalogues are presented in \citet{Reidetal:2015}. The samples jointly cover a large cosmic volume ($V_{\rm eff}=7.4\,{\rm Gpc}^3$) with a number density of galaxies that ensures that the shot noise does not dominate at BAO scales at  the relevant redshifts. Obviously in the edges of the redshift-bin of the LOWZ and CMASS samples the shot noise has a more relevant role than close to the center of the redshift-bin, just because of the difference in the number density of galaxies. However,  the power spectrum signal is dominated by those galaxies of the center of the bin, which is less affected by shot noise. Full details of the catalogues are provided in \citet{Reidetal:2015}.

In order to correct for several observational artefacts in the catalogues, the CMASS and LOWZ samples incorporate a set of weights: a redshift failure weight, $w_{\rm rf}$, a fibre collision weight, $w_{\rm fc}$, and a systematic weight, $w_{\rm sys}$ (CMASS only), which combines a seeing condition weight and a stellar weight \citep{Rossetal:2012,Andersonetal:2014,Reidetal:2015}. Hence, each galaxy target contributes to our estimate of the target galaxy density field by
\begin{equation}
\label{eq:wc}w_c=w_{\rm sys}(w_{\rm rf}+w_{\rm fc}-1).
\end{equation}
The redshift failure weights account for galaxies that have been observed, but whose redshifts have not been measured: nearby galaxies, which are approximated as being ``equivalent'', are up-weighted to remove any bias in the resulting field. The fibre collision weight similarly corrects for galaxies that could not be observed as there was another target within $62''$, a physical limitation of the spectrograph (see \cite{Rossetal:2012} for details). The systematic weight accounts for fluctuations in the target density caused by changes in the observational efficiency. This effect is only present for the CMASS sample, which relies on deeper imaging data;  such a weight is not required for the brighter LOWZ sample \citep{Tojeiro14}.

Additionally, we adopt the standard weight to balance regions of high and low density \citep{FKP,Beutleretal:2013},
\begin{equation}
\label{eq:wfkp}w_{\rm FKP}({\bf r})=\frac{w_{\rm sys}({\bf r})}{w_{\rm sys}({\bf r})+w_c({\bf r})n({\bf r})P_0},
\end{equation}
where $n$ is the mean number density of galaxies and $P_0$ is the amplitude of the galaxy power spectrum at the scale where the error is minimised. We assume $P_0=10\,000\,h^{-3}\,{\rm Mpc}^3$, which corresponds to the amplitude of the power spectrum at scales $k\sim0.10\,h\,{\rm Mpc}^{-1}$ \citep{Reidetal:2015}.

\subsection{The mock survey catalogues}\label{sec:mocks}

Mock samples are a key component in the analysis of precision cosmological data provided by galaxy surveys. They are a fundamental requirement  to test and analyse the large-scale structure and they help determine systematic errors on the measurements. Most of the relevant large-scale physics is captured by approximate methods, so we do not necessarily need to base mock catalogues on full N-body cosmological simulations; small numbers of N-body simulations can instead be used to calibrate a more efficient scheme. In this paper we use mocks created by two different approaches: MultiDark-Patchy BOSS DR12 mocks\footnote{http://data.sdss3.org/datamodel/index-files.html} (hereafter \textsc{MD-Patchy} mocks) \citep[][companion paper]{Kitauraetal2015} and Quick-Particle-Mesh mocks \citep{QPMmocks}, hereafter \textsc{qpm} mocks. Both schemes incorporate observational effects including the survey selection window, veto mask and fiber collisions.  In Paper I we demonstrated that the covariance matrices obtained using these two types of mocks were  similar and yielded  similar errors-bars and cross covariance parameters (see Fig. 3 and 13 in Paper I).  Because of this, in this paper we perform two parallel analyses of the data using \textsc{MD-Patchy} and \textsc{qpm} mocks. Since the measurements and the errors coming from the two sets of mocks are very similar, we average both the measurements and errors in order to obtain a single value for each parameter.
Since the covariance matrix is estimated from a set of mocks, its inverse  is biased due to the limited number of realizations. We account for this effect by applying the correction proposed by \cite{Hartlap07}. In addition to this scaling, we have to propagate the error in the covariance matrix to the error on the estimated parameters. We can do this by scaling the variance for each parameter by the factor of eq. 18 of \cite{Percival13}. The corrections to the error bars described above are small  ($< 2\%$ for \textsc{qpm} and $< 1\%$ for \textsc{MD-Patchy}), but are included in all the error bars quoted in this paper.

\subsection{Fiducial Cosmology}\label{sec:fiducial_cosmo}

In this paper we analyse the data assuming a fiducial cosmological model. The values of this cosmological model, as well as the cosmologies of \textsc{qpm} and \textsc{MD-Patchy} mocks, are presented in Table \ref{table:cosmo}.

\begin{table}
\begin{center}
\begin{tabular}{|c|c|ccccc}
 & $\Omega_m$ & $\Omega_\Lambda$ & $\Omega_bh^2$ & $h$ & $n_s$ \\
 \hline
 \hline
 Fid. & 0.31 & 0.69 & 0.023569 & 0.7 &  0.9624 \\
 \textsc{MD-Patchy} & 0.307115 & 0.692885 & 0.022045 & 0.6777 & 0.96 \\
 \textsc{qpm} & 0.29 & 0.71 & 0.022442 & 0.7 & 0.97 \\
\end{tabular}
\end{center}
\caption{Cosmological parameters chosen for the fiducial cosmology, for the \textsc{MD-Patchy} cosmology and \textsc{qpm} cosmology. }
\label{table:cosmo}
\end{table}%

 In \S\ref{sec:test_mocks} we show that the arbitrary choice of cosmology has a systematic effect of $\lesssim 0.3\%$ on the peak position values obtained from the \textsc{qpm} mocks. 
 We have checked the effect of this systematic error by adding it in quadrature along with the statistical errors for the data.  We have found that this represents just a 5\% increase of the size of the total error-bars budget for the case of the position of the BAO peak in the isotropic post-recon signal in the CMASS sample, and much less in the rest of the variables. Because of this we consider this effect as sub-dominant and we will not consider it in the results of this paper. We present a further discussion of this point later in \S\ref{sec:test_mocks}. 
 
\subsection{Reconstruction}\label{sec:postrecon}

The initial BAO signal in the clustering is damped by the comoving motions of galaxies, potentially reducing the fidelity of BAO measurements. Using the observed density field, we can model the galaxy motions, and move the over-density back to its ``original'' position, recovering a fraction of this signal \citep{Eis07}. The fraction of the signal recovered is dependent on the density of galaxies, which limits how well the displacement field can be determined \citep{Burden1}. Redshift-Space Distortions (RSD) complicate the implementation of this method, as they produce a LOS-dependent distortion that has to be included when estimating the displacement field. In Fourier-space, an iterative method can be used to compensate for this effect \citep{burden2}, where a correction is computed at each step using the prior estimate of displacements. An alternative approach is to grid the galaxy over-density field, and determine RSD and displacements using a finite-difference routine based on the values of the over-densities at grid-points \citep{Padmanabhan08}. Both methods provide consistent results;  we use the finite-difference method here. We will make measurements both before applying this algorithm (pre-recon), and after (post-recon).

\section{Power Spectrum Estimator}\label{sec:estimator_ps}

We  build the galaxy power spectrum multipole estimator by defining the function
\begin{equation}
F({\bf r})=\frac{w_{\rm FKP}({\bf r})}{I_2^{1/2}} [w_c({\bf r})n({\bf r}) - \alpha_{\rm ran} n_s({\bf r})],
\label{eq:FKP_factor}
\end{equation}
where $n$ and $n_s$ are, respectively, the observed number density of galaxies and the number density of a synthetic catalog Poisson sampled with the same mask and selection function as the survey with no other cosmological correlations. The functions $w_c$ and $w_{\rm FKP}$ were defined in Eqs.~(\ref{eq:wc}) and~(\ref{eq:wfkp}) respectively. The factor $\alpha_{\rm ran}$ is the ratio between the weighted number of observed galaxies over the random catalogue, $\alpha_{\rm ran}\equiv \sum_i^{N_{\rm gal}} w_c / N_s$, where $N_s$ denotes the number of objects in the synthetic catalog and $N_{\rm gal}$ the number of galaxies in the real catalog. The factor $I_2$ normalises the amplitude of the observed power in accordance with its definition in a universe with no survey selection. 

We compute the power spectrum multipoles using the implementation of the Yamamoto estimator \citep{Yamamotoetal:2006} presented in \cite{Bianchietal:2015}, which is based on using multiple Fast Fourier Transforms, keeping the relevant LOS information by approximating the LOS of each pair of galaxies with the LOS of one of the two galaxies (see section 3 of Paper I for more details on the algorithm used). This is a reliable approximation on the scales of interest, which clearly improves on assuming a single fixed LOS for the whole survey for the quadrupole, but will eventually break down at  large scales and high order multipoles \citep{Yooetal:2015,Samushiaetal:2015}.

We use a random catalogue of number density of $\bar{n}_s({\bf r})=\alpha_{\rm ran}^{-1} \bar{n}({\bf r})$, with $\alpha_{\rm ran}^{-1}\simeq50$, for pre-recon, and $\alpha_{\rm ran}^{-1}\simeq20$ for post-recon catalogues. We place the LOWZ and CMASS galaxy samples on $1024^3$ quadrangular grids, of a box of side $L_b=2300\,h^{-1}{\rm Mpc}$ for the LOWZ galaxies, and $L_b=3500\,h^{-1}{\rm Mpc}$  to fit the CMASS galaxies. This approach corresponds to a grid-cell resolution of $3.42\,h^{-1}{\rm Mpc}$ for the CMASS galaxies and $2.25\,h^{-1}{\rm Mpc}$ for the LOWZ galaxies. The fundamental wave-lengths are therefore $k_f=1.795\cdot10^{-3}\,h\,{\rm Mpc}^{-1}$ and $k_f=2.732\cdot10^{-3}\,h\,{\rm Mpc}^{-1}$ for the CMASS and LOWZ galaxies, respectively. We apply the Cloud-in-Cells scheme (CiC) to associate galaxies to grid-cells, and bin the power spectrum $k-$modes in 60 bins between the fundamental frequency $k_f$ and a maximum frequency of $k_{\rm M}=0.33\,h\,{\rm Mpc}$,  with width  $\Delta \log_{10} k=\left[ \log_{10}(k_{\rm M})-\log_{10}(k_f) \right]/60$.

We limit the scales fitted as follows: our procedure for determining the largest scale for the fitting process is based on limiting the impact of the systematic weights, and is presented in appendix A of Paper I. We limit scales to $k>0.02\,h\,{\rm Mpc}^{-1}$ for the monopole and $k>0.04\,h\,{\rm Mpc}^{-1}$ for the quadrupole. The smaller scale used for the BAO peak determination is $k_{\rm max}=0.3\,h\,{\rm Mpc}^{-1}$, as  for smaller scales the BAO information is quite limited. 

In this paper we work with the combined north and south samples,  NGC+SGC,  for all the power spectrum multipoles. This combination is performed by averaging both NGC and SGC components weighted by their area, $P_{\rm NGC+SGC}=(P_{\rm NGC} A_{\rm NGC} + P_{\rm SGC} A_{\rm SGC})/(A_{\rm NGC}+A_{\rm SGC})$, both for LOWZ and CMASS samples. The values of the areas are: $A_{\rm NGC}^{\rm LOWZ}=5836\,{\rm deg}^2$, $A_{\rm SGC}^{\rm LOWZ}=2501\,{\rm deg}^2$, $A_{\rm NGC}^{\rm CMASS}=6851\,{\rm deg}^2$ and $A_{\rm SGC}^{\rm CMASS}=2525\,{\rm deg}^2$.

\section{Fitting the power spectrum moments}\label{sec:methodology}

\subsection{Modelling the power spectrum moments}

We model the power spectrum multipoles in order to measure the position of the BAO features and marginalise over the broad-band information. To reduce the complexity of the fits, we choose to fit the monopole and the $\mu^2$-moment, defined as $P^{(\mu^2)}\equiv2/5P^{(2)}(k)+P^{(0)}(k)$, rather than the monopole and quadrupole. As the pair monopole - $\mu^2$-moment are a linear transformation of the pair monopole - quadrupole, the information content is the same in both. $P^{(\mu^2)}$ has the property that, in linear theory, the shape is the same as $P^{(0)}$, and we can therefore use the same modelling procedure for both, albeit with different free parameters describing the deviations from linear theory to account for LOS dependent effects.
 
To create the model moments, we start by computing the  linear power spectrum at a given cosmology, $P_{\rm lin}$, which  we generate using \textsc{Camb} \citep{Lewisetal:2000}. The linear power spectrum was separated into two components, one containing the BAO oscillations, $O_{\rm lin}$, and a smooth component, $P_{\rm lin, sm}$; $P_{\rm lin}(k)=O_{\rm lin}(k)P_{\rm lin,sm}(k)$. This separation was performed using the same method applied to the data, but employing an analytic model for the BAO and smooth model based on the fitting formulae of \citet{EH98}.
The position of the BAO peak is described by  $\alpha$, which parametrises the features of the oscillations as a function of $k$ in the $O_{\rm lin}$-function,  $O_{\rm lin}(k/\alpha)$. We use a superscript ``(0)'' for the monopole and ``(2)'' for the value measured from the $\mu^2$-moment. We adopt the model described in \citet{Andersonetal:2014} to account for deviations in the smooth fit away from this linear model. The full model is
\begin{equation}
  P_{\rm model}(k;\alpha)=P_{\rm model,sm}(k)\left\{ 1+[O_{\rm lin}(k/\alpha)-1]e^{-\frac{1}{2}k^2{\Sigma}^2_{\rm nl}} \right\},
\label{Pmodel}
\end{equation}
where $P_{\rm model,sm}$ is a phenomenological parametrisation of the non-linear power spectrum monopole with no-BAO,
\begin{equation}
  P_{\rm model,sm}(k)=B^2 P_{\rm lin, sm}(k) +A_1 k+ A_2 + 
    \frac{A_3}{k}+\frac{A_4}{k^2}+\frac{A_5}{k^3},
\label{eq:Psmooth}
\end{equation}
and  $B$, $A_i$ and $\Sigma_{\rm nl}$ are free parameters that account for redshift space distortions and nonlinearities. Together with $\alpha$,  there are 8 free parameters for each model. These parameters are allowed to be different in the fits to the monopole and $\mu^2$ moment and we adopt superscripts ``(0)'' for the monopole and ``(2)'' for the fit to the $\mu^2$-moment, where appropriate. 

The survey window acts as a convolution of the power spectrum moments, smoothing the true power spectrum to produce that observed. As the survey window affects the BAO, it cannot be described by a variation of the free parameters described above. To include the window effects, we follow the approach described in \S5.4 of Paper I, producing matrices to account for these effects through a discrete convolution
\begin{eqnarray}
\nonumber P^{(0)}_{\rm win.}(k_i)&=&\sum_j \mathcal{W}_{ij}^{00} {P^{(0)}}_{\rm model}(k_j) + \sum_j \mathcal{W}_{ij}^{02} {P^{(2)}}_{\rm model}(k_j),\\
\nonumber P^{(2)}_{\rm win.}(k_i)&=&\sum_j \mathcal{W}_{ij}^{20} {P^{(0)}}_{\rm model}(k_j) + \sum_j \mathcal{W}_{ij}^{22} {P^{(2)}}_{\rm model}(k_j),\\
\label{eq:Wijll}
\end{eqnarray}
where, $\mathcal{W}_{ij}^{nn'}$ are the elements of the window-survey matrix, ${P^{(n)}}_{\rm model}$ are the pre-masked models described by Eq.~(\ref{Pmodel}), and $P^{(i)}_{\rm win.}$ are the final products for the observed moments. Details of how the $\mathcal{W}_{ij}^{nn'}$ elements are estimated can be found in eq. (14) of Paper I. This model is able to fit both pre- and post-recon power spectrum moments with residuals that are of far lower significance than the BAO signal (see \S\ref{sec:test_mocks}).

\subsection{Finding the best-fitting parameters} \label{sec:mcmc}

The monopole and $\mu^2$-moment are assumed to be drawn from a multi-variate Gaussian distribution, whose covariance can be computed using the galaxy mocks described in \S\ref{sec:mocks}. For the pre-recon analysis we consider two different sets of mocks, \textsc{qpm} and \textsc{MD-Patchy}, whereas for the post-recon analysis we only use \textsc{qpm} mocks. For details about the computation of the covariance matrices we refer the reader to section 6.1 of Paper I. 

Eq.~(\ref{Pmodel}) includes 8 free parameters for each of the two power spectrum moments, so we have 16 free parameters in total to be fitted to each set of measurements. To investigate this parameter space we use a bespoke Markov-Chain Monte-Carlo (MCMC) routine, run for each fit for $10^8$ steps split into 10 sub-chains, which satisfies the convergence check that each sub-chain gives consistent results. Each chain was started close to the best-fitting locations found using a downhill simplex routine, and a burn-in of $10^5$ steps was removed from each. To reduce the computational burden, each chain was thinned by a factor $10^2$ before measuring marginalised parameters and errors. We only consider the interval $0.8<\alpha_i<1.2$ for $\alpha_0$ and $\alpha_2$ to avoid unphysical solutions and to reduce the convergence time of the MCMC routine. This is a very conservative prior which is never hit by the MCMC routine, and that happens to be at more than $5\sigma$ of the final results. Due to noise, secondary maxima where the BAO signal is located within the large-scale, noisy, part of the power spectrum, can cause a mobility issue for the chains. Given that the data storage and analysis are computationally expensive, we choose to run relatively long chains, which are less affected by the secondary maxima, and thin them to allow the subsequent analysis to be performed quickly.

 The correlation among the parameters of interest, $\alpha_0$ and $\alpha_2$ is shown later in Fig. \ref{plot:scatter_data}. The correlation among the $\alpha$-parameters and the rest of nuisance parameters, $B$, $A_i$ and $\Sigma_{\rm nl}$, is found to be very weak or consistent with 0 ($|r|<0.10$). The strongest is a weak correlation of $r\simeq0.27$ between $\alpha_0$ and $\Sigma_{\rm nl}$; and between $\alpha_2$ and $\Sigma_{\rm nl}$ for the CMASS sample; and $r\simeq-0.30$ between  $\alpha_0$ and $\Sigma_{\rm nl}$; and between $\alpha_2$ and $\Sigma_{\rm nl}$ for the LOWZ sample.

\subsection{Interpreting the measured BAO scales}\label{sec:AP}

Our BAO scale measurements are given by $\alpha_0$ and $\alpha_2$, and their covariance. To translate these parameters into easily-to-model cosmological measurements, we adopt the standard definitions of the Alcock-Paczynski (AP) parameters,  
\begin{equation}
  \alpha_{\parallel}\equiv\frac{H^{\rm fid}(z) r_s^{\rm fid}(z_d)}{H(z) r_s(z_d)};\quad     
  \alpha_{\perp}\equiv\frac{D_A(z)r_s^{\rm fid}(z_d)}{D_A^{\rm fid}(z)r_s(z_d)}, 
\label{a_def}
\end{equation}
Our measurements of $\alpha_0$ and $\alpha_2$ each provide degenerate measurement of $\alpha_{||}$ and
$\alpha_{\perp}$. \citet{Ross15} showed that we should expect a degeneracy of the kind,
\begin{equation}  
  \alpha_{i}^{m+n} = \alpha_{||}^m \alpha_{\perp}^n,
  \label{aarelation}
\end{equation}
where for the post-recon monopole (and RSD removal), $m=1/3$ and $n=2/3$, and for the $\mu^2$-moment, $m=3/5$ and $n=2/5$. Our window function convolution accounts for any deviations from these ideal solutions caused by the survey geometry. Thus, the AP parameters, $\alpha_\parallel$ and $\alpha_\perp$ can be written in terms of the position of the BAO peak in the monopole and $\mu^2$-moment,
\begin{equation}
\alpha_\parallel=\alpha_0^{-3/2}\alpha_2^{5/2};\quad\alpha_\perp=\alpha_0^{9/4}\alpha_2^{-5/4}.
\end{equation}
Using these expressions we relate the BAO peak position in the monopole and $\mu^2$-moment to the Hubble parameter and the angular distance as,
\begin{equation}
H(z) r_s(z_d)=[H(z) r_s(z_d)]^{\rm fid}\alpha_0^{3/2}\alpha_2^{-5/2}
\end{equation}
and
\begin{equation}
D_A(z)/r_s(z_d)=[D_A(z)/r_s(z_d)]^{\rm fid}\alpha_0^{9/4}\alpha_2^{-5/4}.
\end{equation}
For the fiducial cosmological model of Table \ref{table:cosmo}, $[H(z_{\rm LOWZ})r_s(z_d)]^{\rm fid}=11.914\cdot10^3\,{\rm km}s^{-1}$ and $[D_A(z_{\rm LOWZ})/r_s(z_d)]^{\rm fid}=6.667$ for the LOWZ sample and $[H(z_{\rm CMASS})r_s(z_d)]^{\rm fid}=13.827\cdot10^3\,{\rm km}s^{-1}$ and $[D_A(z_{\rm CMASS})/r_s(z_d)]^{\rm fid}=9.330$ for the LOWZ sample. In this fiducial cosmological model we have $r_s(z_d)=143.70\,{\rm Mpc}$. We report these parameters and its correlation coefficient as a final result.

\section{Tests on galaxy mocks}\label{sec:test_mocks}

In this section we test for potential systematics in the BAO model as well as those associated to the fitting algorithm. We focus on \textsc{qpm} post-recon mocks, which we  analyse assuming using two different cosmologies (with $\Omega_m=0.29$ or $\Omega_m=0.31$) and two different BAO models (with $\Omega_bh^2=0.022442$ or $\Omega_b h^2=0.023569$). For the four different combinations of $\Omega_m$ and $\Omega_b h^2$, we test that we are able to recover the expected BAO shifts positions even when the cosmology or the BAO model assumed for analysing the mocks does not necessarily  match their true values. In the case where both the cosmology and the BAO model assumed match those values from the mocks, the expected values for $\alpha_0$ and $\alpha_2$ are 1. However, when either the cosmology or the BAO model assumed differ from those of the mocks, we expect that the values for $\alpha_0$ and $\alpha_2$ will differ from 1, with the expected values given by Eqs.~(\ref{a_def}). We will refer to these expected BAO peak position as $\alpha_0^{\rm exp}$ and $\alpha_2^{\rm exp}$. 

\begin{table*}
\begin{center}
\begin{tabular}{|c|c|c|c|c|c|cc}
Statistic & Sample & BAO model & Cosmology & $\alpha_0-\alpha_0^{\rm exp}$ & $\alpha_0^{\rm exp}$ & $\alpha_2-\alpha_2^{\rm exp}$ & $\alpha_2^{\rm exp}$\\
\hline
\hline 
\multirow{8}{*}{$ \alpha(\widetilde{P})$} & \multirow{4}{*}{  LOWZ } & \multirow{2}{*}{$\Omega_bh^2=0.023569$}  & $\Omega_m=0.29$ & $0.00137 \pm 0.00074$ & $0.976687$ & $0.0022 \pm0.0010$ &  $0.976687$\\  & & & $\Omega_m=0.31$ & $0.00061 \pm0.00048$ & $0.982765$ & $0.00103 \pm0.00066$ & $0.984004$\\ \cline{3-8} &  & \multirow{2}{*}{$\Omega_bh^2=0.022442$} &$\Omega_m=0.29$ & $0.00137 \pm0.00048$ & $1.000000$ & $0.00175 \pm0.00068$ & $1.000000$\\ & &  & $\Omega_m=0.31$ & $0.00131 \pm0.00050$ & $1.006223$ & $0.00158 \pm0.00069$ & $1.007491$\\ \cline{2-8}&  \multirow{4}{*}{  CMASS } & \multirow{2}{*}{$\Omega_bh^2=0.023569$}  & $\Omega_m=0.29$ & $-0.00253 \pm0.00031$ & $0.976687$ & $-0.00257 \pm0.00044$ &  $0.976687$ \\& & & $\Omega_m=0.31$ & $-0.00274 \pm0.00031$ & $0.986808$ & $-0.00290 \pm0.00044$ & $0.988829$\\ \cline{3-8} &  & \multirow{2}{*}{$\Omega_bh^2=0.022442$} &$\Omega_m=0.29$ & $-0.00098 \pm0.00032$ & $1.000000$ & $-0.00102 \pm0.00044$ & $1.000000$\\ & &  & $\Omega_m=0.31$ & $-0.00127 \pm0.00032$ & $1.010363$ & $-0.00147 \pm0.00045$ & $1.012431$  \\

\hline
\multirow{8}{*}{$ \widetilde{\alpha}({P})$} & \multirow{4}{*}{  LOWZ } & \multirow{2}{*}{$\Omega_bh^2=0.023569$}  & $\Omega_m=0.29$ & $0.00173 \pm 0.00050$ & $0.976687$ & $0.00227 \pm0.00071$ &  $0.976687$ \\  & & & $\Omega_m=0.31$ & $0.00204 \pm 0.00051$ & $0.982765$ & $0.00254 \pm 0.00073$ & $0.984004$ \\ \cline{3-8} &  & \multirow{2}{*}{$\Omega_bh^2=0.022442$} &$\Omega_m=0.29$ & $0.00153 \pm 0.00051$ & $1.000000$ & $0.00140 \pm 0.00072$ & $1.000000$ \\ & &  & $\Omega_m=0.31$ & $0.00177 \pm 0.00052$ & $1.006223$ & $0.00155 \pm 0.00073$ & $1.007491$ \\ \cline{2-8}&  \multirow{4}{*}{  CMASS } & \multirow{2}{*}{$\Omega_bh^2=0.023569$}  & $\Omega_m=0.29$ & $-0.00243 \pm 0.00032$ & $0.976687$ & $-0.00244 \pm 0.00045$ &  $0.976687$ \\ & & & $\Omega_m=0.31$ & $-0.00224 \pm 0.00033$ & $0.986808$ & $-0.00209 \pm 0.00045$ & $0.988829$ \\ \cline{3-8} &  & \multirow{2}{*}{$\Omega_bh^2=0.022442$} &$\Omega_m=0.29$ & $-0.00135 \pm 0.00033$ & $1.000000$ & $-0.00147 \pm 0.00045$ & $1.000000$ \\ & &  & $\Omega_m=0.31$ & $-0.00112\pm0.00034$ & $1.010363$ & $-0.00116\pm0.00046$ & $1.012431$  \\
\end{tabular}
\caption{Measured BAO peak positions $\alpha_0$ and $\alpha_2$ recovered from the mocks and presented in terms of their expected values. These results have been obtained from the  \textsc{qpm} post-recon mocks for the LOWZ and CMASS samples, for different assumptions of cosmologies and different BAO models. We show the difference between the measured and the expected BAO position in the monopole, $\alpha_0-\alpha_0^{\rm exp}$ and in the $\mu^2$-moment, $\alpha_2-\alpha_2^{\rm exp}$. For reference the expected BAO peak position for each cosmology and BAO model for both moments are listed. Two different cases are analysed: $\alpha(\widetilde P)$, where the BAO peak position is obtained from the fit to the mean of 1000 mocks; and $\widetilde{\alpha}(P)$, where the BAO peak position is obtained from the mean of the BAO peak position fits to each individual mock. The former statistic is only sensitive to the systematic errors of the model, whereas the latter includes noise-dependent effects. }
\label{table:systematics}
\end{center}
\end{table*}%

We start by computing the best-fitting values of $\alpha_0$ and $\alpha_2$ for the average monopole and $\mu^2$-moment over the 1000 realisations of the mocks, for the four different combination of $\Omega_m$ and $\Omega_b h^2$ (labelled $\alpha(\widetilde{P})$ in Table~\ref{table:systematics}). By averaging the power spectrum moment over the realisations we reduce the noise-dependent systematic errors to a negligible level compared to the systematic errors of the model. Therefore, when we compare the obtained $\alpha_i$ with its expected value we are exclusively testing the potential systematics of the BAO model itself. Given the results of Table \ref{table:systematics} associated to the statistic $\alpha(\widetilde{P})$  we can state that the BAO model is able to reproduce the expected value of the BAO peak position in both the monopole and  $\mu^2$-moment with $<0.30\%$ accuracy. 

In order to test for noise-dependent systematic errors we have also computed expected value for $\alpha_0$ and $\alpha_2$ for each mock and average among them.  Unlike the former case, now we do not cancel the noise-dependent systematic associated to each individual fit.  The results are listed in Table \ref{table:systematics}, labelled $\widetilde{\alpha}(P)$ (i.e. the average of the BAO peak position of the mocks). For some values of the cosmology and BAO model we observe that the deviation between the expected and the recovered values has significantly changed with respect to fitting the average power from the mocks, $\alpha(\widetilde{P})$. This result is  due to the remaining statistical noise associated to each individual mock fit. In general,  we are able to recover the BAO peak position in both the monopole and $\mu^2$-moment with $\leq0.25\%$ accuracy. These values are significantly below the statistical errors on the measurements, in a similar way as we have seen for the $\alpha(\widetilde{P})$ statistic. We therefore conclude that the potential systematic errors, both noise-dependent and model-dependent, do not play any important role in the BAO peak position estimation in the monopole or $\mu^2$-moment, where the statistical errors of the survey dominate. We therefore do not apply any correction to the obtained results from the  data.

\section{Results}\label{sec:results}

\begin{figure*} \centering \includegraphics[scale=0.45]{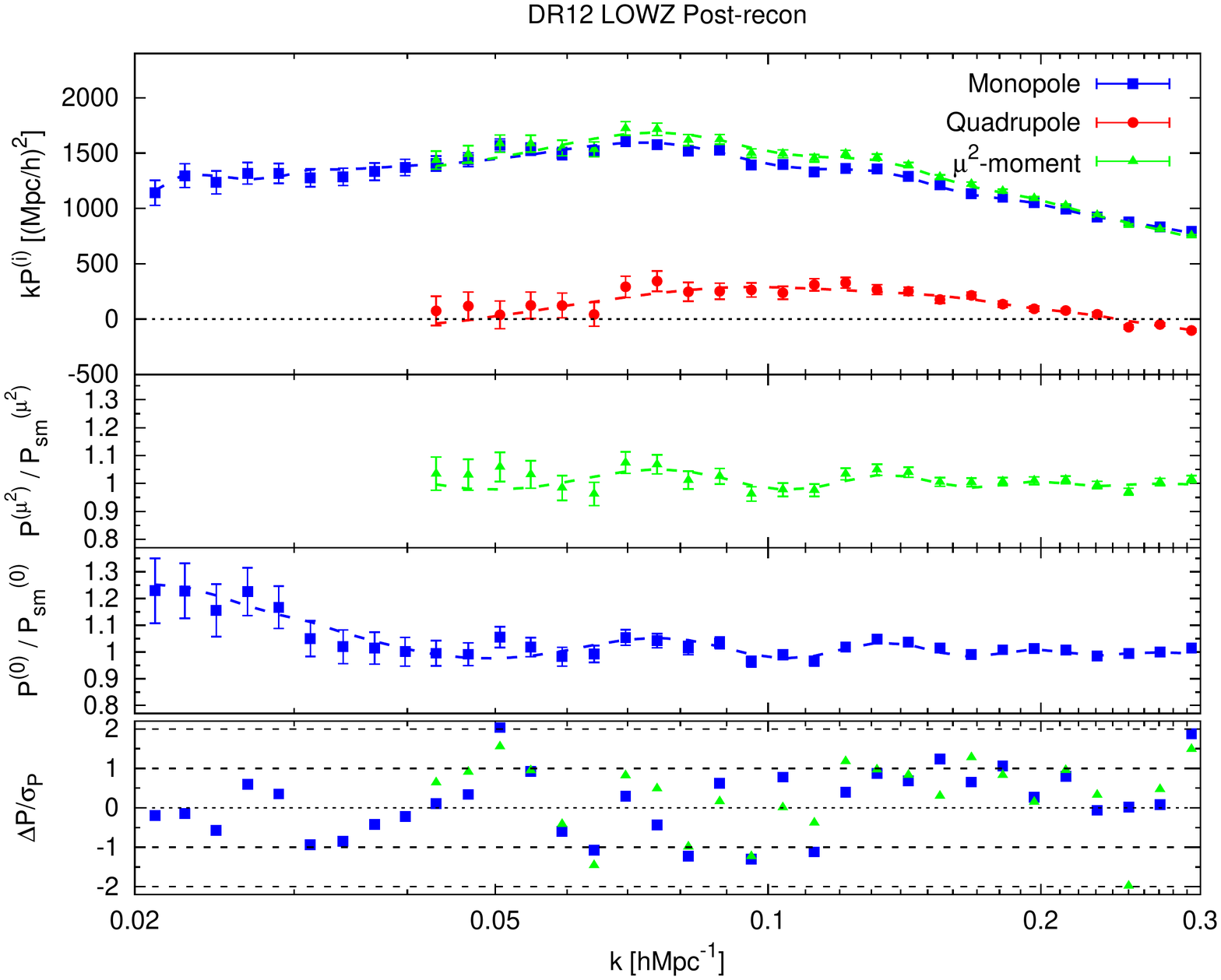} \includegraphics[scale=0.45]{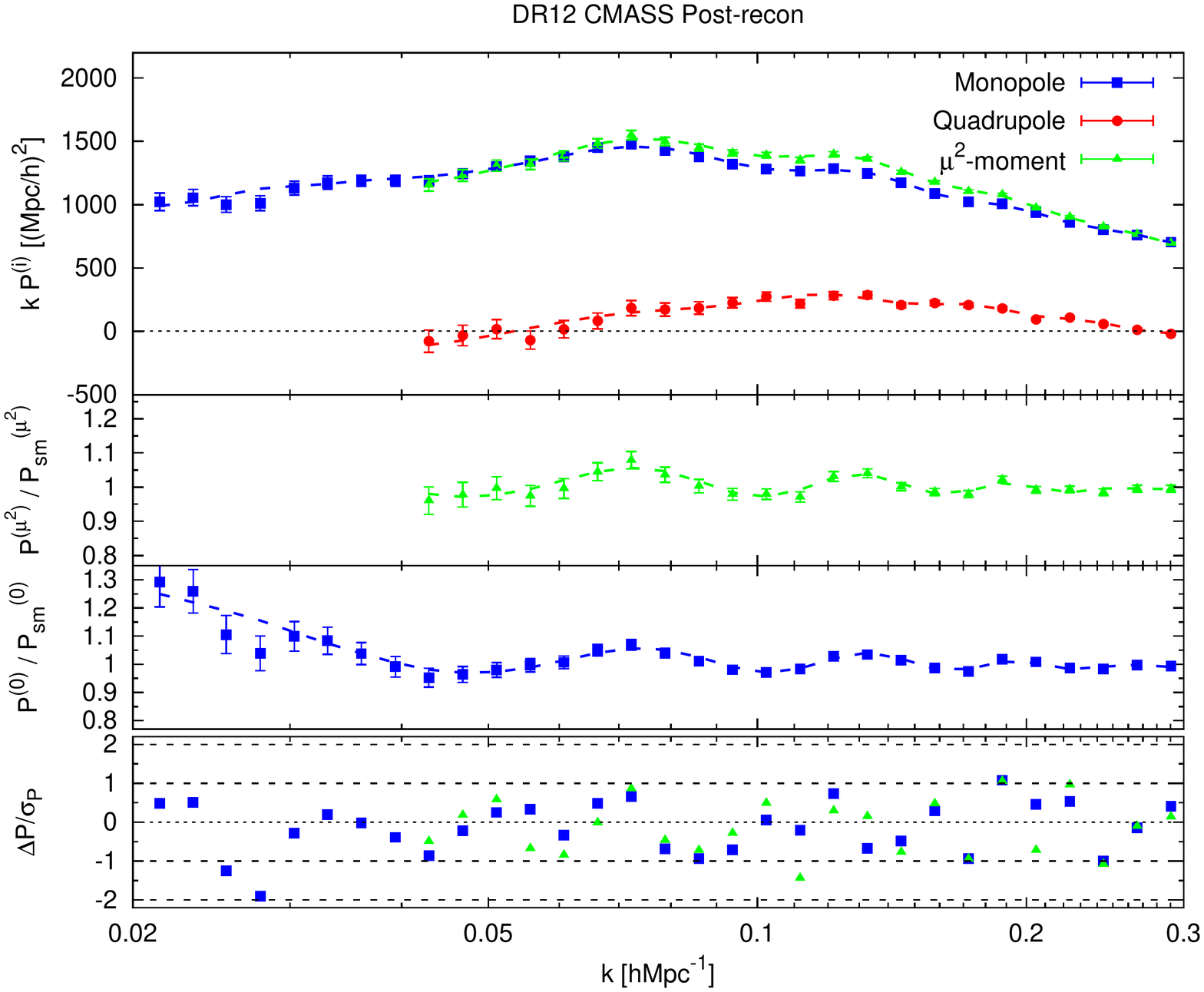} 
  \caption{The measured LOWZ (top panel) and CMASS (bottom panel) DR12 post-recon, monopole (blue squares), quadrupole (red circles) and $\mu^2$-moment (green triangles) power spectra. For all the cases the measurements correspond to a combination of the northern and southern galaxy caps according to their effective areas as described in \S\ref{sec:estimator_ps}. The error-bars are calculated from the dispersion of measurements  using the \textsc{qpm} mocks. The red, blue and green lines correspond to the best-fitting model of Eq.~(\ref{Pmodel}) with the BAO peak position as a free parameter. Within each panel we also present the power spectrum monopole and $\mu^2$-moment divided by the smooth power spectrum calculated in our fit to the data. For the monopole and $\mu^2$-moment we see how the model is able to capture the BAO features observed in the data. For clarity, the bottom sub-panels show the residuals between the measurement and the model, $\Delta P\equiv P^{\rm model} - P^{\rm data}$, divided by the $1\sigma$ error of the data. The black dashed lines marks the $1\sigma$ and $2\sigma$ deviations.}
\label{plot:model_data}
\end{figure*}

Fig.~\ref{plot:model_data} presents the post-recon power spectrum monopole (blue squares), quadrupole (blue circles) and $\mu^2$-moment (green triangles), for the LOWZ and CMASS samples of the DR12 data measurements (top and bottom panels as labeled) from the combination of the NGC and SCG. The coloured lines show the best-fitting model given by Eq.~(\ref{Pmodel}) with the BAO peak position for the monopole ($\alpha_0$) and for the $\mu^2$-moment ($\alpha_2$). For simplicity we only show the best-fitting model according to the covariance matrix extracted from the \textsc{qpm} mocks. We will later show in Table \ref{table:data_results} the results from the data assuming both \textsc{qpm} and \textsc{MD-Patchy} covariance. The model for the quadrupole has been obtained from the models for the monopole and $\mu^2$-moment. As the BAO peak positions in the monopole and $\mu^2$-moment are  similar to those in the fiducial model, there is no apparent BAO signature in the quadrupole. Within each panel, we also display the data and best-fitting model for the monopole and $\mu^2$-moment divided by the smoothed power spectrum $P_{\rm sm}$ of Eq.~(\ref{eq:Psmooth}) for the best-fitting model. For clarity, the bottom sub-panels show the residuals between the measurement and the model, $\Delta P\equiv P^{\rm model} - P^{\rm data}$, divided by the $1\sigma$ error of the data. The black dashed lines marks the $1\sigma$ and $2\sigma$ deviations. The model is able to accurately reproduce  the BAO features both in the monopole and $\mu^2$-moment for both LOWZ and CMASS sample. As expected, the BAO features are more significant in the monopole than in the $\mu^2$-moment, as the former has a higher signal-to-noise ratio. Similarly, the BAO features are stronger in the CMASS sample than in the LOWZ.

At large scales, the quadrupole measured for both LOWZ and CMASS samples is small, consistent with the reconstruction process removing the linear component of the RSD. However, the reconstruction process is not able to fully suppress the whole RSD signal and non-linear components are left, as is the AP effect caused by the potential differences between the true underlying $\Omega_m$ and the fiducial value assumed, in this case $\Omega_m=0.31$. 

\begin{figure}
\centering
\includegraphics[scale=0.3]{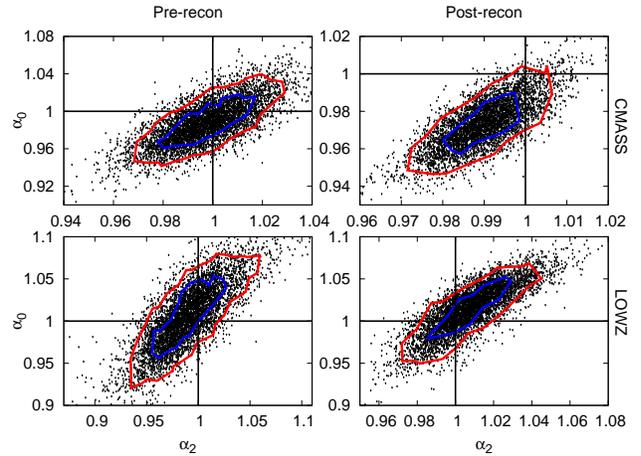}
\caption{Likelihood surfaces in the $\alpha_0-\alpha_2$ parameter space obtained from measuring the power spectrum DR12 data monopole and $\mu^2$-moment. The black points are the (down-sampled) output of the MCMC chain, for the CMASS and LOWZ samples, for both pre- and post-reconstructed catalogues, as labeled. In all cases the \textsc{qpm}-covariance matrix has been used for estimating the power spectrum errors and their correlations. The blue and red contours show the 68\% and 95.4\% confident regions, respectively.}
\label{plot:scatter_data}
\end{figure}

 Fig.~\ref{plot:scatter_data} displays  the output of the MCMC chains when estimating the BAO peak position in the monopole and $\mu^2$-moment, in the pre- and post-recon data catalogues, for the LOWZ and CMASS samples, as labeled in each panel. The blue and red contours display the 68\% and 95.4\% confident regions, respectively In this case the \textsc{qpm} covariance matrix has been used in all cases. By comparing the pre- and post-recon panels we see how the likelihood surface in the $\alpha_0-\alpha_2$ parameter space is significantly reduced in the both LOWZ and CMASS sample, both for $\alpha_0$ and $\alpha_2$. These results are consistent with DR11 results \citep{Andersonetal:2014,Tojeiro14}.

 Table~\ref{table:data_results} presents the marginalised measurements of $\alpha_0$ and $\alpha_2$ (and the derived AP parameters, $\alpha_\parallel$ and $\alpha_\perp$ for the post-recon samples) calculated from the MCMC chains (such as those presented in Fig.~\ref{plot:scatter_data}), for the following cases: 
\begin{enumerate}
\item Pre-recon catalogues when the \textsc{MD-Patchy} covariance matrix has been used. 
\item Pre-recon catalogues adopting  the \textsc{qpm} covariance matrix. 
\item Post-recon catalogues when the \textsc{MD-Patchy} covariance matrix has been used. 
\item Post-recon catalogues when the \textsc{qpm} covariance matrix has been used. 
\end{enumerate}

\begin{table*}
\begin{center}
\begin{tabular}{c|c|c|c|c|cc|c|c|c}
Sample & Catalogue &  Covariance & $\alpha_0$ & $\alpha_2$ & $r_{02}$ & $\alpha_\parallel$ & $\alpha_\perp$ & $r_{\parallel\perp}$ & $\chi^2/{\rm dof}$ \\
 \hline
 \hline
 \multirow{3}{*}{LOWZ} &   \multirow{2}{*}{{Pre-recon}} & \textsc{MD-Patchy} & $0.998\pm0.028$ & $1.019\pm0.037$ & $0.78$ & $1.053\pm0.067$  & $0.974\pm0.039$ & $-0.33$ & $37.16/43$ \\ &  &  \textsc{qpm} & $0.989\pm 0.031$ & $1.005\pm0.039$ & $0.78$ & $1.031\pm0.069$ &  $0.971\pm0.042$ & $-0.32$ & $35.72/43$ \\ \cline{2-10} &   \multirow{2}{*}{{Post-recon}} & \textsc{MD-Patchy} & $1.009\pm0.017$ & $1.018\pm0.027$ & $0.81$ & $1.032\pm0.050$  & $0.999\pm0.023$ & $-0.40$ & $59.53/43$ \\ &  &  \textsc{qpm} & $1.009\pm0.019$ & $1.016\pm0.029$ & $0.81$ & $1.027\pm0.053$ &  $1.001\pm0.025$ & $-0.42$ & $55.57/43$ \\
 \hline
  \multirow{3}{*}{CMASS} &   \multirow{2}{*}{{Pre-recon}} & \textsc{MD-Patchy} & $1.002\pm0.015$ & $0.996\pm0.024$ & $0.75$ & $0.987\pm0.045$ & $1.010\pm0.023$ &  $-0.52$& $43.73/38$  \\ &  & \textsc{qpm} & $0.997\pm0.016$ & $ 0.988\pm0.024$ & $0.73$ & $0.974\pm0.045$ & $1.010\pm0.025$ & $-0.52$ & $38.48/38$  \\ \cline{2-10}& \multirow{2}{*}{{Post-recon}} & \textsc{MD-Patchy} & $0.9895\pm0.0091$ & $0.974\pm0.013$ & $0.75$ & $0.950\pm0.024$ & $1.010\pm0.014$ &  $ -0.43$& $49.00/38$  \\ &  & \textsc{qpm} & $0.9899\pm0.0088$ & $ 0.974\pm0.014$ & $0.74$ & $0.950\pm0.025$ & $1.010\pm0.014$ & $-0.51$ & $37.24/38$

\end{tabular}
\caption{Marginalised measurements of the BAO peak position for the monopole, $\alpha_0$, and for the $\mu^2$-moment,  $\alpha_2$, in the LOWZ and CMASS sample of the DR12 data as indicated. Also listed are their cross-correlation coefficient, $r_{02}$. For the post-recon samples we also show the derived AP parameters, $\alpha_\parallel$ and $\alpha_\perp$ with their corresponding cross-correlation coefficient, $r_{\parallel\perp}$. We include the value of the best-fitting $\chi^2$ divided by the number of degrees of freedom (dof). The different rows are for the pre and post-recon catalogues, when both \textsc{qpm} and \textsc{MD-Patchy} covariances are used.}
\label{table:data_results}
\end{center}
\end{table*}

The pre-recon results from both \textsc{MD-Patchy} and \textsc{qpm} covariance matrices agree very well, both in the measurement and errors, suggesting that, as it was observed for the RSD analysis in Paper I, the impact of the differences between mocks when translated into the covariance matrices of the power spectra is not significant. For pre-recon, we measure the BAO peak position in the monopole with $\simeq3\%$ and $\simeq1.5\%$ precision in the LOWZ and CMASS samples, respectively. In the $\mu^2$-moment we are able to measure the BAO peak position with $\simeq4\%$ and $\simeq2.4\%$ precision in the LOWZ and CMASS samples, respectively. 
 
 The post-recon results using either \textsc{MD-Patchy} or \textsc{qpm} covariance matrices also agree very well, with differences that are much smaller than the uncertainties due to the statistical errors. The post-recon results show an improvement on the level of precision for the BAO peak position determination, both in the monopole and in the $\mu^2$-moment. We are able to improve the determination of the BAO peak position in the monopole down to $\sim1.8\%$ and $\sim0.90\%$ precision for the LOWZ and CMASS samples, respectively. The BAO peak position in the $\mu^2$-moment is determined with $\sim2.8\%$ and $\sim1.4\%$ precision for the LOWZ and CAMSS samples, respectively. For the CMASS sample, this result represents an improvement of $\simeq50\%$, both in $\alpha_0$ and $\alpha_2$. For the LOWZ sample the improvement is of $\simeq40\%$ for $\alpha_0$ and $\sim25\%$ for $\alpha_2$. 

\begin{figure*}
\centering
\includegraphics[trim = 0mm 25mm 20mm 0mm, clip=false, scale=0.34]{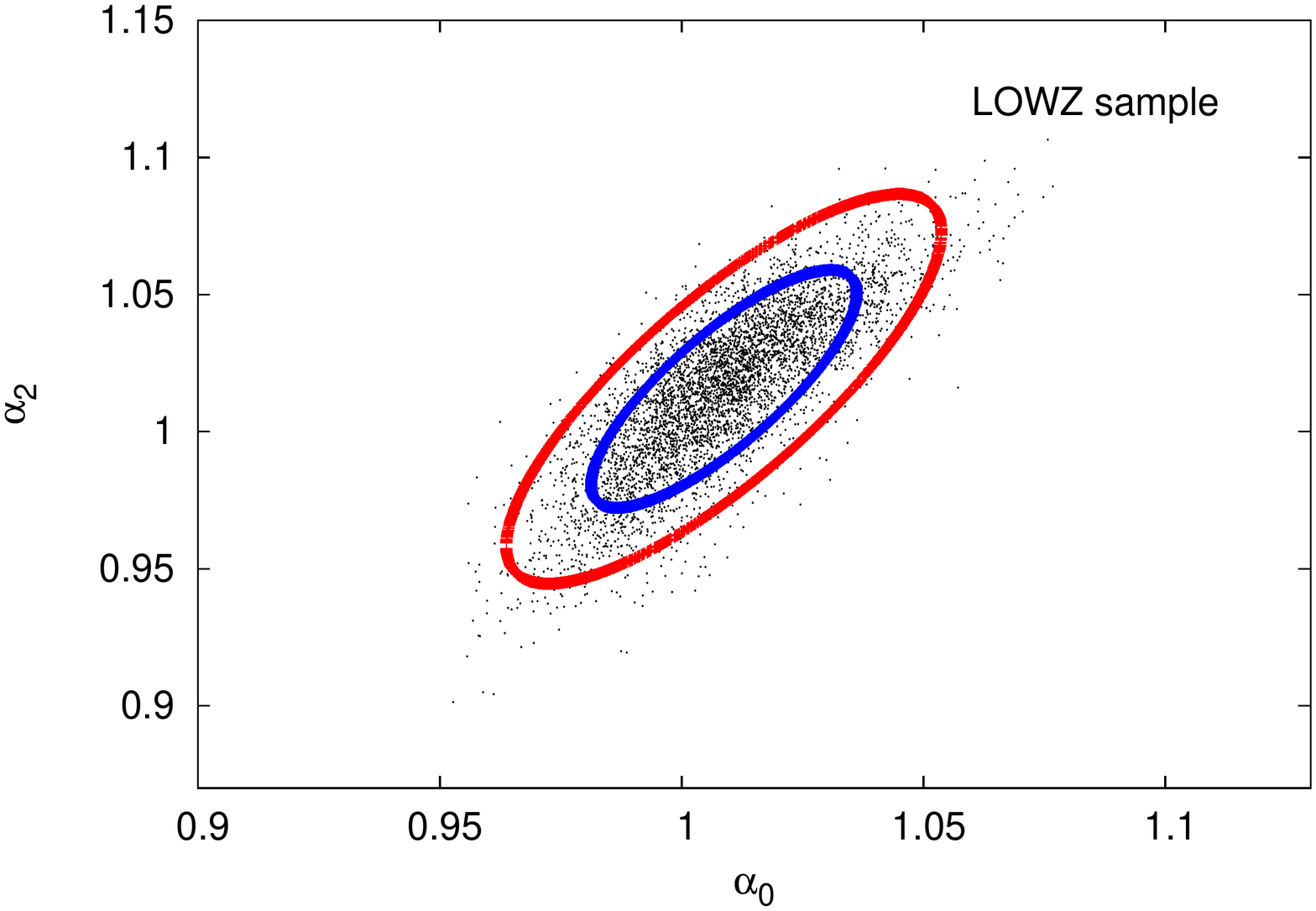}
\includegraphics[trim = 20mm 25mm 0mm 0mm, clip=false,scale=0.34]{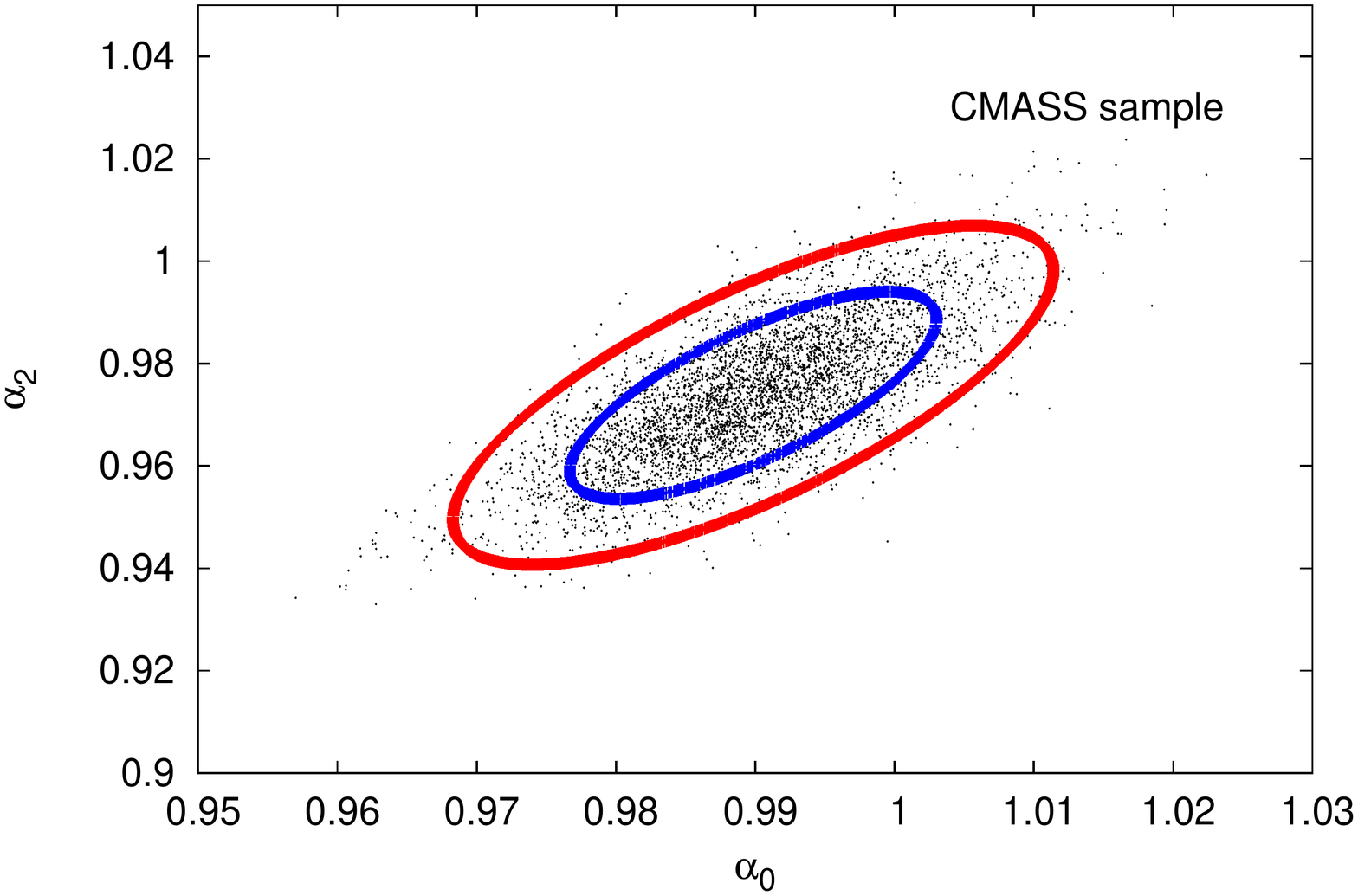}

\includegraphics[trim = 0mm 20mm 20mm 25mm, clip=false, scale=0.34]{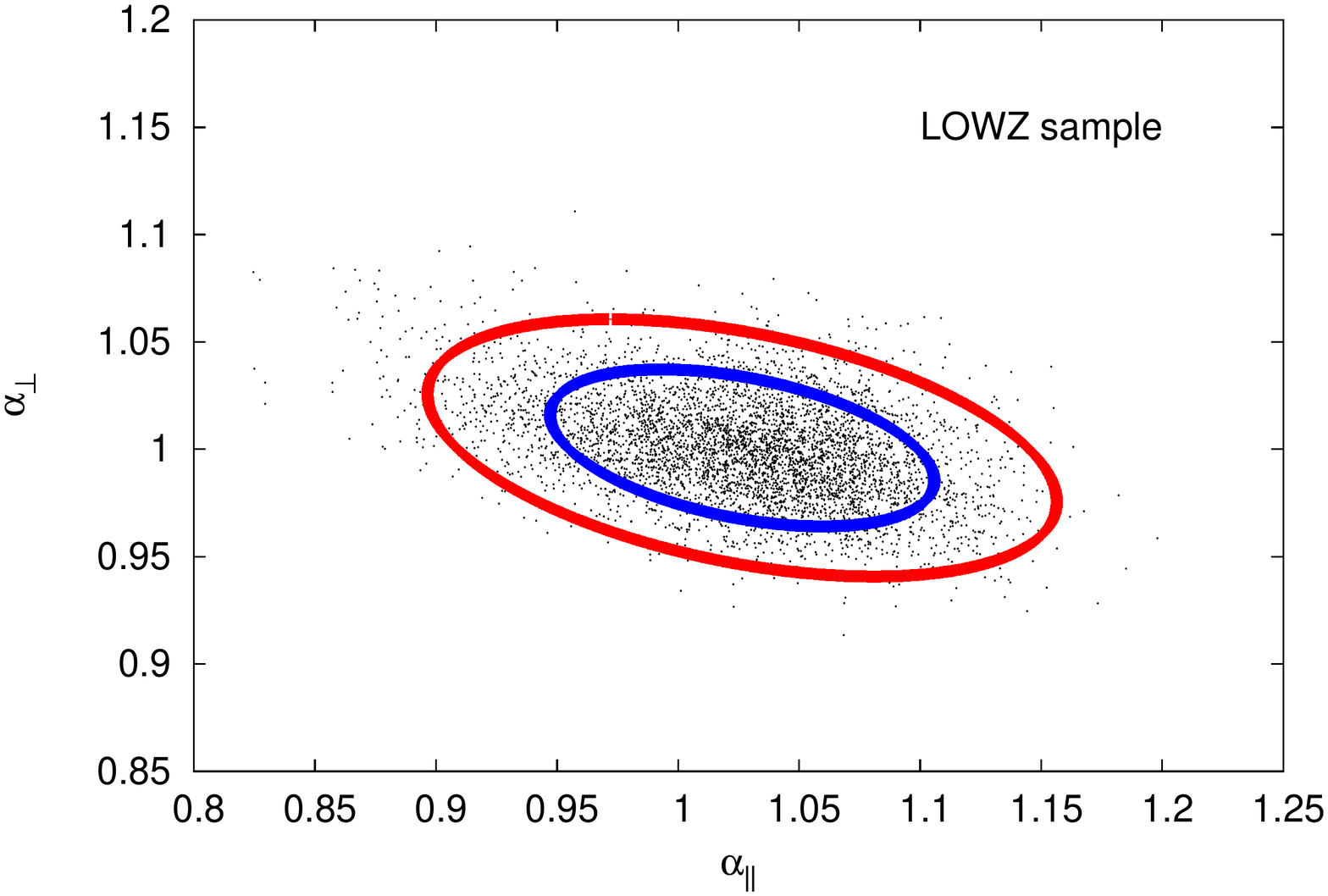}
\includegraphics[trim = 20mm 20mm 0mm 25mm, clip=false, scale=0.34]{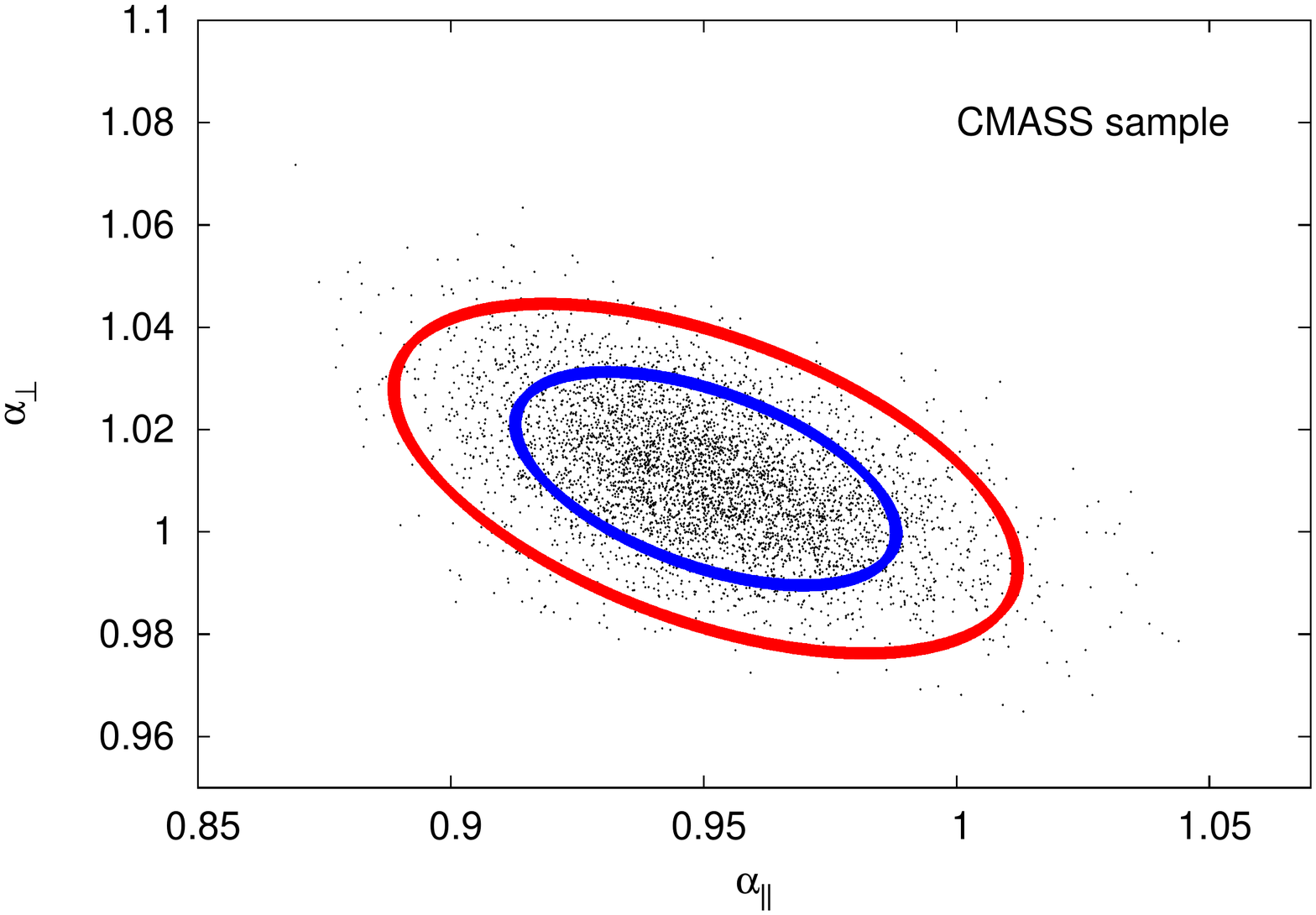}

\caption{Distribution of the MCMC points for the DR12 data post-recon catalogues where the \textsc{qpm} covariance matrix is used, in terms of the BAO peak position variables ($\alpha_0$ and $\alpha_2$, in the top panels) and the AP parameters ($\alpha_\parallel$ and $\alpha_\perp$, in the bottom panels) for the LOWZ sample (left panels) and for the CMASS sample (right panels). In blue lines the $1\sigma$ ellipses ($\Delta\chi^2=2.30$) and in red lines the $2\sigma$ ellipses ($\Delta\chi^2=6.17$) corresponding to Gaussian fits to the likelihood based on the parameters given Table~\ref{table:data_results}. For all cases the distribution of points is close to a Gaussian distribution.}
\label{plot:gauss}
\end{figure*}

The top panels of Fig.~\ref{plot:gauss} show the distribution of points in our MCMC chains for the post-recon LOWZ and CMASS DR12 data samples. We have included $1\sigma$ ($\Delta\chi^2=2.30$ in blue ellipses) and $2\sigma$ ($\Delta\chi^2=6.17$ in red ellipses) contours from the Gaussian fit corresponding to the parameters extracted from the \textsc{qpm} covariance quoted in Table~\ref{table:data_results}. For the BAO peak position variables, $\alpha_0$ and $\alpha_2$, as well as the AP variables $\alpha_\parallel$ and $\alpha_\perp$, the  distribution of points  matches well the Gaussian fits. 

Combining the best-fitting results on $\alpha_\parallel$ and $\alpha_\perp$ of Table \ref{table:data_results} for both \textsc{qpm} and \textsc{MD-Patchy}-based covariance matrices and the relations of Eq. (\ref{a_def}), we report, $H(z_{\rm LOWZ})r_s(z_d)=(11.60\pm0.60)\cdot10^3\,{\rm km}s^{-1}$ and $D_A(z_{\rm LOWZ})/r_s(z_d)=6.66\pm0.16$ for the LOWZ sample with a cross correlation coefficient of $r_{HD_A}=0.41$; and $H(z_{\rm CMASS})r_s(z_d)=(14.56\pm0.37)\cdot10^3\,{\rm km}s^{-1}$ and $D_A(z_{\rm CMASS})/r_s(z_d)=9.42\pm0.13$ with a cross correlation coefficient of $r_{HD_A}=0.47$. 
These quantities represent for $H(z)r_s(z_d)$ a 5.2\% and a 2.5\% measurement precision, for LOWZ and CMASS, respectively; and for $D_A(z)/r_s(z_d)$ a 2.4\% and a 1.4\% measurement precision, for LOWZ and CMASS, respectively.  We can also put constraints on $D_V(z)=(cz(1+z)D_A^2H^{-1})^{1/3}$, which is the combination of parameters upon which the BAO location would depend for a galaxy sample with an isotropically distributed set of equally weighted pair separations. We find that for LOWZ $D_V(z_{\rm LOWZ})/r_s(z_d)=8.62\pm0.15$; whereas for CMASS $D_V(z_{\rm CMASS})/r_s(z_d)=13.70\pm0.12$.

\subsection{Consensus Values}

\begin{table*}
\begin{center}
\begin{tabular}{c|c|c|c|c|c}
Sample & Statistic & $H(z)r_s(z_d)\,[10^3\,{\rm km}s^{-1}]$ & $D_A(z)/r_s(z_d)$ & $r_{HD_A}$ & $D_V(z)/r_s(z_d)$  \\
 \hline
 \hline
 \multirow{3}{*}{LOWZ} &   Power Spectrum & $11.60\pm0.60$ & $6.66\pm0.16$  & $0.41$ & $8.62\pm0.15$ \\  &  Correlation Function & $11.65\pm0.81$ & $6.67\pm0.14$ & $0.29$ & $8.59\pm0.15$ \\ \cline{2-6} & Consensus & $11.63\pm 0.69$ & $6.67 \pm 0.15$ & $0.35$ & $8.61\pm0.15$ \\
\hline
\multirow{3}{*}{CMASS} &   Power Spectrum & $14.56\pm0.37$ & $9.42\pm0.13$  & $0.47$ & $13.70\pm0.12$ \\  &  Correlation Function & $14.75\pm0.50$ & $9.52\pm0.13$ & $0.57$ & $13.79\pm0.14$ \\ \cline{2-6} & Consensus & $14.67\pm 0.42$ & $9.47 \pm0.12$ & $0.52$ & $13.74\pm0.13$ \\
\end{tabular}
\caption{Distance constrain parameters, $H(z)r_s(z_d)$, $D_A(z)/r_s(z_d)$ and their correlation coefficient $r_{HD_A}$,  inferred from the post-recon power spectrum (this work) and correlation function analysis \citep[][companion paper]{Cuestaetal2015}, and the corresponding consensus value. The correlation coefficient between the power spectrum and the correlation function has been determined using the \textsc{qpm} post-recon mocks.  In addition, we show the derived parameter $D_V$.} 
\label{table:consensus}
\end{center}
\end{table*}

We combine the measurements presented above with those obtained using the correlation function monopole, $\xi^{(0)}$, and quadrupole, $\xi^{(2)}$, of the same dataset reported in \citep[][companion paper]{Cuestaetal2015}, and report the consensus  values for $H(z)r_s(z_d)$, $D_A(z)/r_s(z_d)$\footnote{Note that in \citep[][companion paper]{Cuestaetal2015} a different fiducial cosmology has been used to compute $\alpha_\parallel$ and $\alpha_\perp$. However, the results in terms of $H(z)r_s(z_d)$ and $D_A(z)/r_s(z_d)$ should not depend on this assumption.} and $D_V(z)/r_s(z_d)$. We summarise these values in Table \ref{table:consensus}. We see that both analyses agree well within $1\sigma$ error-bars. Although in theory the power spectrum and the correlation function contain the same information, in practice the analysis is performed on a finite range of scales, and the information content of the two estimators differs slightly. The correlation between them is large, as reported in the DR11analysis \citep{Andersonetal:2014} with a correlation factor of $r\simeq0.95$ for $\alpha_0$. Using the post-recon \textsc{qpm} mocks we have determined that for DR12 data the correlation factor between the power spectrum and the correlation function is 0.91 in the LOWZ and 0.83 in the CMASS sample, for both $\alpha_\parallel$ and $\alpha_\perp$. This change is not produced by the differences between the geometries of DR11 and DR12, which are very similar, but because for DR11 the fitting to the isotropic correlation function was performed using 8 bin-positions, whereas for the anisotropic correlation function of DR12 was performed only using 1 bin-position. This makes the DR12 anisotropic correlation function fits noisier with respect to the DR11 ones, but at the same time, the correlation factor between $\xi$ and $P$ is reduced. In this way, the reduction on the error-bars produced by using more bin-positions is partially canceled by the fact that the correlation factor approaches to 1.  Thus, the final results on the consensus value of the error-bars does not strongly depend on the number of bin-positions used in the correlation function analysis. 

From Table \ref{table:consensus} we also note that both for LOWZ and CMASS the errors on $H(z)r_s(z_d)$ are $\sim25\%$ smaller when this quantity is estimated from the power spectrum. On the other hand, the errors on $D_A(z)/r_s(z_d)$ are similar for both LOWZ and CMASS. We have observed these differences not only from the likelihood of the data, but from the dispersion of the mocks, which demonstrates that the origin of this discrepancy is related to the methodology of how $P^{(0)}(k)$ and $P^{(\mu^2)}$ are fit compared to how $\xi^{(0)}$ and $\xi^{(2)}$ are, and not with the data. These differences suggest that fitting the BAO peak position in the $\mu^2$-moment (instead of fitting the quadrupole) is able to constrain better the AP distortion in the radial direction, both for LOWZ and CMASS.   

Using this correlation factor, we combine the measurements to obtain a consensus value,
\begin{eqnarray}
\alpha_{\rm con}&=&\frac{\alpha_P+\alpha_{\xi}}{2},\\
\sigma_{\rm con}&=&\frac{\sigma_P+\sigma_{\xi}}{2}\left( \frac{1+r}{2}  \right)^{1/2},
\end{eqnarray}
where $\alpha_P\pm\sigma_P$ and $\alpha_{\xi}\pm\sigma_{\xi}$ are the AP parameters measured from the power spectrum and correlation fraction, respectively, and $\alpha_{\rm con}\pm\sigma_{\rm con}$ the consensus value. When $r=0$, both measurements would be independent and we would get a reduction of a factor of 2 on the error-bars, whereas when $r=1$ both measurements are fully correlated and therefore the error-bars is unchanged. Using the correlation factors found in this paper, combining the correlation function and power-spectrum measurements improve the error bars by a few percent, both in LOWZ and CMASS.

\section{Conclusions}\label{sec:conclusions}

We have presented a new method to use Baryon Acoustic Oscillations, measured with respect to the line-of-sight (LOS), to make cosmological measurements. The method fits the model of \citet{Andersonetal:2014} to both the monopole and $\mu^2$-moments, providing two correlated measurements of BAO positions using both statistics. This technique allows the clean separation of the BAO component from the broad-band power for both spherically-averaged determinations and measurements that are anisotropic around the LOS. This separation  is harder to achieve when fitting the quadrupole where the BAO signal is removed if the fiducial cosmology matches the true one. We have applied this method to measurements of the anisotropic power spectra for the LOWZ and CMASS DR12 galaxies of the Baryon Oscillations Spectroscopic Survey of the Sloan Digital Sky Survey III, presented in Paper I, to constrain the BAO peak position in the monopole ($\alpha_0$) and in the $\mu^2$-moment ($\alpha_2$). 

We have tested potential systematics of the BAO model using the post-recon \textsc{qpm} mocks catalogues, assuming different cosmology parameters and different BAO features models. We  are able to constrain the BAO peak position in the monopole and $\mu^2$-moment  better than $0.3\%$ accuracy. These values are several times smaller than the $1\sigma$ statistical errors in the LOWZ and CMASS samples. Furthermore, the scatter in these values appears random, and a significant component is likely due to residual noise. These tests explicitly demonstrate that the method is independent of the fiducial cosmology assumed when calculating the power spectrum, or the BAO model to be fitted to the data.

We have used the \textsc{MD-Patchy} and \textsc{qpm} mocks in the pre-reconstructed and  post-reconstructed catalogues to estimate the covariance matrices and hence the error-bars in the BAO peak position parameters. We have found no significant differences in the results. From the post-reconstructed DR12 data power spectrum monopole and $\mu^2$-moment we have measured: $\alpha_0(z_{\rm LOWZ})=1.009\pm0.018$ and $\alpha_2(z_{\rm LOWZ})=1.017\pm0.028$ with $r_{02}=0.81$, where $z_{\rm LOWZ}=0.32$; and $\alpha_0(z_{\rm CMASS})=0.9897\pm0.0090$ and $\alpha_2({\rm z_{CMASS}})=0.974\pm0.014$ with $r_{02}=0.75$, where $z_{\rm CMASS}=0.57$. We report these measurements in terms of cosmological parameters: $H(z_{\rm LOWZ})r_s(z_d)=(11.60\pm0.60)\cdot10^3\,{\rm km}s^{-1}$ and $D_A(z_{\rm LOWZ})/r_s(z_d)=6.66\pm0.16$ with a cross-correlation coefficient of $r_{HD_A}=0.41$, for the LOWZ sample; and $H(z_{\rm CMASS})r_s(z_d)=(14.56\pm0.37)\cdot10^3\,{\rm km}s^{-1}$ and $D_A(z_{\rm CMASS})/r_s(z_d)=9.42\pm0.13$ with a cross-correlation coefficient of $r_{HD_A}=0.47$, for the CMASS sample.

We have combined these measurements with those obtained using the correlation function monopole, $\xi^{(0)}$, and quadrupole, $\xi^{(2)}$, of the same dataset reported in \citep[][companion paper]{Cuestaetal2015}. We report $H(z_{\rm LOWZ})r_s(z_d)=(11.63\pm0.69)\cdot10^3\,{\rm km}s^{-1}$ and $D_A(z_{\rm LOWZ})/r_s(z_d)=6.67\pm0.15$ with a cross-correlation coefficient of $r_{HD_A}=0.35$, for the LOWZ sample; and $H(z_{\rm CMASS})r_s(z_d)=(14.67\pm0.42)\cdot10^3\,{\rm km}s^{-1}$ and $D_A(z_{\rm CMASS})/r_s(z_d)=9.47\pm0.12$ with a cross-correlation coefficient of $r_{HD_A}=0.52$, for the CMASS sample. We see that the results reported from the analysis of the power spectrum agree well within $1\sigma$ error-bars with the consensus values inferred from combining the power spectrum and 2-point correlation function analyses.

\section*{Acknowledgements}
HGM is grateful for support from the UK Science and Technology Facilities Council through the grant
ST/I001204/1.
WJP is grateful for support from the UK Science and Technology Facilities Research Council through the grant
ST/I001204/1, and the European  Research Council through the grant ``Darksurvey", reference 614030.
FSK acknowledges the support of the Karl-Schwarzschild Program from the Leibniz Society.

Funding for SDSS-III has been provided by the Alfred P. Sloan
Foundation, the Participating Institutions, the National Science
Foundation, and the U.S. Department of Energy Office of Science. The
SDSS-III web site is http://www.sdss3.org/.

SDSS-III is managed by the Astrophysical Research Consortium for the
Participating Institutions of the SDSS-III Collaboration including the
University of Arizona,
the Brazilian Participation Group,
Brookhaven National Laboratory,
University of Cambridge,
Carnegie Mellon University,
University of Florida,
the French Participation Group,
the German Participation Group,
Harvard University,
the Instituto de Astrofisica de Canarias,
the Michigan State/Notre Dame/JINA Participation Group,
Johns Hopkins University,
Lawrence Berkeley National Laboratory,
Max Planck Institute for Astrophysics,
Max Planck Institute for Extraterrestrial Physics,
New Mexico State University,
New York University,
Ohio State University,
Pennsylvania State University,
University of Portsmouth,
Princeton University,
the Spanish Participation Group,
University of Tokyo,
University of Utah,
Vanderbilt University,
University of Virginia,
University of Washington,
and Yale University.
This research used resources of the National Energy Research Scientific
Computing Center, which is supported by the Office of Science of the
U.S. Department of Energy under Contract No. DE-AC02-05CH11231.

Numerical computations were done on the Sciama High Performance Compute (HPC) cluster which is supported by the ICG, SEPNet and the University of Portsmouth.

%
%
%


\def\jnl@style{\it}
\def\aaref@jnl#1{{\jnl@style#1}}

\def\aaref@jnl#1{{\jnl@style#1}}

\def\aj{\aaref@jnl{AJ}}                   
\def\araa{\aaref@jnl{ARA\&A}}             
\def\apj{\aaref@jnl{ApJ}}                 
\def\apjl{\aaref@jnl{ApJ}}                
\def\apjs{\aaref@jnl{ApJS}}               
\def\ao{\aaref@jnl{Appl.~Opt.}}           
\def\apss{\aaref@jnl{Ap\&SS}}             
\def\aap{\aaref@jnl{A\&A}}                
\def\aapr{\aaref@jnl{A\&A~Rev.}}          
\def\aaps{\aaref@jnl{A\&AS}}              
\def\azh{\aaref@jnl{AZh}}                 
\def\baas{\aaref@jnl{BAAS}}               
\def\jrasc{\aaref@jnl{JRASC}}             
\def\memras{\aaref@jnl{MmRAS}}            
\def\mnras{\aaref@jnl{MNRAS}}             
\def\pra{\aaref@jnl{Phys.~Rev.~A}}        
\def\prb{\aaref@jnl{Phys.~Rev.~B}}        
\def\prc{\aaref@jnl{Phys.~Rev.~C}}        
\def\prd{\aaref@jnl{Phys.~Rev.~D}}        
\def\pre{\aaref@jnl{Phys.~Rev.~E}}        
\def\prl{\aaref@jnl{Phys.~Rev.~Lett.}}    
\def\pasp{\aaref@jnl{PASP}}               
\def\pasj{\aaref@jnl{PASJ}}               
\def\qjras{\aaref@jnl{QJRAS}}             
\def\skytel{\aaref@jnl{S\&T}}             
\def\solphys{\aaref@jnl{Sol.~Phys.}}      
\def\sovast{\aaref@jnl{Soviet~Ast.}}      
\def\ssr{\aaref@jnl{Space~Sci.~Rev.}}     
\def\zap{\aaref@jnl{ZAp}}                 
\def\nat{\aaref@jnl{Nature}}              
\def\iaucirc{\aaref@jnl{IAU~Circ.}}       
\def\aplett{\aaref@jnl{Astrophys.~Lett.}} 
\def\apspr{\aaref@jnl{Astrophys.~Space~Phys.~Res.}}
\def\bain{\aaref@jnl{Bull.~Astron.~Inst.~Netherlands}} 
\def\fcp{\aaref@jnl{Fund.~Cosmic~Phys.}}  
\def\gca{\aaref@jnl{Geochim.~Cosmochim.~Acta}}   
\def\grl{\aaref@jnl{Geophys.~Res.~Lett.}} 
\def\jcp{\aaref@jnl{J.~Chem.~Phys.}}      
\def\jgr{\aaref@jnl{J.~Geophys.~Res.}}    
\def\jqsrt{\aaref@jnl{J.~Quant.~Spec.~Radiat.~Transf.}}
\def\memsai{\aaref@jnl{Mem.~Soc.~Astron.~Italiana}}
\def\nphysa{\aaref@jnl{Nucl.~Phys.~A}}   
\def\physrep{\aaref@jnl{Phys.~Rep.}}   
\def\physscr{\aaref@jnl{Phys.~Scr}}   
\def\planss{\aaref@jnl{Planet.~Space~Sci.}}   
\def\procspie{\aaref@jnl{Proc.~SPIE}}   
\def\jcap{\aaref@jnl{J. Cosmology Astropart. Phys.}}

\let\astap=\aap
\let\apjlett=\apjl
\let\apjsupp=\apjs
\let\applopt=\ao

\newcommand{\etal}{et al.\ }

\newcommand{\mpc}{\, {\rm Mpc}}
\newcommand{\kpc}{\, {\rm kpc}}
\newcommand{\hmpc}{\, h^{-1} \mpc}
\newcommand{\ihmpc}{\, h\, {\rm Mpc}^{-1}}
\newcommand{\ikms}{\, {\rm s\, km}^{-1}}
\newcommand{\kms}{\, {\rm km\, s}^{-1}}
\newcommand{\hkpc}{\, h^{-1} \kpc}
\newcommand{\lya}{Ly$\alpha$\ }
\newcommand{\lyb}{Lyman-$\beta$\ }
\newcommand{\lyaf}{Ly$\alpha$ forest}
\newcommand{\lr}{\lambda_{{\rm rest}}}
\newcommand{\bF}{\bar{F}}
\newcommand{\bS}{\bar{S}}
\newcommand{\bC}{\bar{C}}
\newcommand{\bB}{\bar{B}}
\newcommand{\vdF}{{\mathbf \delta_F}}
\newcommand{\vdS}{{\mathbf \delta_S}}
\newcommand{\vdf}{{\mathbf \delta_f}}
\newcommand{\vdn}{{\mathbf \delta_n}}
\newcommand{\vdC}{{\mathbf \delta_C}}
\newcommand{\vdX}{{\mathbf \delta_X}}
\newcommand{\xrei}{x_{rei}}
\newcommand{\lrmin}{\lambda_{{\rm rest, min}}}
\newcommand{\lrmax}{\lambda_{{\rm rest, max}}}
\newcommand{\lmin}{\lambda_{{\rm min}}}
\newcommand{\lmax}{\lambda_{{\rm max}}}
\newcommand{\hi}{\mbox{H\,{\scriptsize I}\ }}
\newcommand{\heii}{\mbox{He\,{\scriptsize II}\ }}
\newcommand{\vp}{\mathbf{p}}
\newcommand{\vq}{\mathbf{q}}
\newcommand{\vxperp}{\mathbf{x_\perp}}
\newcommand{\vkperp}{\mathbf{k_\perp}}
\newcommand{\vrperp}{\mathbf{r_\perp}}
\newcommand{\vx}{\mathbf{x}}
\newcommand{\vy}{\mathbf{y}}
\newcommand{\vk}{\mathbf{k}}
\newcommand{\vR}{\mathbf{r}}
\newcommand{\tdtwo}{\tilde{b}_{\delta^2}}
\newcommand{\tstwo}{\tilde{b}_{s^2}}
\newcommand{\tbthree}{\tilde{b}_3}
\newcommand{\tadtwo}{\tilde{a}_{\delta^2}}
\newcommand{\tastwo}{\tilde{a}_{s^2}}
\newcommand{\tabthree}{\tilde{a}_3}
\newcommand{\vnabla}{\mathbf{\nabla}}
\newcommand{\tpsi}{\tilde{\psi}}
\newcommand{\vv}{\mathbf{v}}
\newcommand{\fnl}{{f_{\rm NL}}}
\newcommand{\tfnl}{{\tilde{f}_{\rm NL}}}
\newcommand{\gnl}{g_{\rm NL}}
\newcommand{\orderfour}{\mathcal{O}\left(\delta_1^4\right)}
\newcommand{\SDSSPF}{\cite{2006ApJS..163...80M}}
\newcommand{\PF}{$P_F^{\rm 1D}(k_\parallel,z)$}
\newcommand\ion[2]{#1$\;${\small \uppercase\expandafter{\romannumeral #2}}}%
\newcommand\ionalt[2]{#1$\;${\scriptsize \uppercase\expandafter{\romannumeral #2}}}%
\newcommand{\vxone}{\mathbf{x_1}}
\newcommand{\vxtwo}{\mathbf{x_2}}
\newcommand{\vRot}{\mathbf{r_{12}}}
\newcommand{\cm}{\, {\rm cm}}

\bibliographystyle{mn2e}
\bibliography{dr12_bao.bib}

   \end{document}